\documentclass[twocolumn,secnumarabic,amssymb,nobibnotes]{revtex4-1}
\usepackage{verbatim}
\usepackage{graphicx}
\usepackage{ifpdf}
\usepackage{tikz}
\usepackage{adjustbox}
\usepackage{geometry}
\usetikzlibrary{shapes,calc}
\usepackage{xcolor}
\pagecolor{white}

\ifpdf
\else
  \usepackage{pst-all}
\fi
\begin{document}
\title{Magnetic, transport, and thermal properties of $\delta$-phase UZr$_2$}%
\author{Xiaxin Ding$^{1}$}%
\email{Xiaxin.Ding@inl.gov}%
\author{Tiankai Yao$^{1}$}%
\author{Lyuwen Fu$^{2}$}%
\author{Zilong Hua$^{1}$}%
\author{Jason Harp$^{1,3}$}%
\author{Chris Marianetti$^{2}$}%
\author{Madhab~Neupane$^{4}$}%
\author{Michael E. Manley$^{3}$}%
\author{David Hurley$^{1}$}%
\author{Krzysztof Gofryk$^{1}$}%
\email{Krzysztof.Gofryk@inl.gov}%
\affiliation{$^{1}$Idaho National Laboratory, Idaho Falls, Idaho 83415, USA\\$^{2}$Columbia University, New York, New York 10027, USA\\$^{3}$Oak Ridge National Laboratory, Oak Ridge, Tennessee 37831, USA\\$^{4}$University of Central Florida, Orlando, Florida 32816, USA}

\date{\today}%

\begin{abstract}
Alloys of hexagonal $\delta$-phase UZr$_2$ have been synthesized and studied by means of heat capacity, magnetic susceptibility, magnetization, electrical resistivity, magnetoresistance, thermoelectric power, thermal conductivity measurements, for the first time, at temperatures from 1.8 to 300 K and in magnetic fields up to 8 T. The weak temperature dependence of the magnetic susceptibility and the small value of both Seebeck (0.75 $\mu$V/K at room temperature) and of the Sommerfeld coefficient (13.5 mJ mol$^{-1}$ K$^{-2}$) point to 5$f$-electrons in this material having a delocalized nature. The electrical resistivity and magnetoresistance indicate the presence of significant electronic disorder in $\delta$-UZr$_2$, consistent with the disorder in its crystal structure. Density functional theory calculations have been performed and compared to experimental results.

\end{abstract}
\maketitle

\section{Introduction}

Actinide systems show a large variety of exotic behavior coming from 5$f$-ligand hybridization. Depending on the strength of this process, unusual behavior has been observed in both actinide metals and alloys~\cite{RevModPhys, RevModPhys.56.755, Sarrao2002, JPSJ.76.063701, PhysRevB.79.134525}. The binary U$X_2$ system in particular is an interesting playground for exploring behavior originating from the dual nature of 5$f$-electrons, and a focus area for search for new electronic phenomena in actinide materials. As shown in Figure 1, the properties of these alloys vary from correlated magnetism, via spin fluctuations, to unconventional superconductivity. For example, USb$_2$ shows both itinerant and localized characters of 5$f$-electrons~\cite{USb2A} with an antiferromagnetic (AFM) ordering below $T_N$ = 203 K~\cite{UP-Bi, PhysRevLett.111.057402}, while UAl$_2$ exhibits delocalized 5$f$-states and spin fluctuations~\cite{UAl2}. More recently, Ran $et~al.$ reported the discovery of spin-triplet superconductivity in UTe$_2$, featuring a transition temperature of 1.6 K and a very large and anisotropic upper critical field exceeding 40~T~\cite{UTe2}. For $\delta$-UZr$_2$, despite some information in the literature on its crystal structure and phase transformations~\cite{Akabori,deltaUZr2,PAUKOV2018113}, its electronic, transport, and magnetic properties are mostly unknown. 

Recently, there has been a renewed interest in the U-Zr system due to its technological importance as a promising metallic nuclear fuel. The intermediate $\delta$ phase is formed on cooling from the high temperature $\gamma$ phase. Akabori $et$ $al$. determined the homogeneity range of the $\delta$ phase, which is 64.2-78.2 at\% Zr at 600 $^{\circ}$C and 66.5-80.2 at\% Zr at 550 $^{\circ}$C~\cite{Akabori}. In the uranium-rich range of the U-Zr phase diagram, transmission electron microscopy (TEM) studies also reveal the coexistence of the $\alpha$ and $\delta$ phases with an alternating lamellar structure~\cite{Urich}. The $\delta$ phase has hexagonal structure with the lattice parameters $a$ = 5.034 $\rm{\AA}$ and $c$ = 3.094 $\rm{\AA}$~\cite{deltaUZr2}. According to high-resolution neutron diffraction measurements~\cite{neutron} and density-functional calculations~\cite{DFC}, the corner (0, 0, 0) sites are occupied solely by the Zr atoms, whereas the two inner positions at ($\frac{2}{3}$, $\frac{1}{3}$, $\frac{1}{2}$) and ($\frac{1}{3}$, $\frac{2}{3}$, $\frac{1}{2}$) are randomly shared by U and Zr atoms, characteristic of disordered structures.

\begin{figure*}[htbp]

\newcommand{\CommonElementTextFormat}[4]
{
  \begin{minipage}{2.2cm}
    \centering
      {\textbf{#1} \hfill #2}%
      \linebreak \linebreak
      {\textbf{#3}}%
      \linebreak \linebreak
      {{#4}}
  \end{minipage}
}

\newcommand{\NaturalElementTextFormat}[4]
{
  \CommonElementTextFormat{#1}{#2}{\huge {#3}}{\large{#4}}
}

\newcommand{\ParaTextFormat}[4]
{
  \CommonElementTextFormat{#1}{#2}{\LARGE {#3}}{\LARGE{#4}}
}

\newcommand{\MagTextFormat}[4]
{
  \CommonElementTextFormat{#1}{\LARGE{#2}}{\LARGE {#3}}{\LARGE{#4}}
}

\newcommand{\OutlineText}[1]
{
\ifpdf
  %
  \pdfliteral direct {0.5 w 1 Tr}{#1}%
  \pdfliteral direct {1 w 0 Tr}%
\else
  %
  \pscharpath[shadow=false,
    fillstyle=solid,
    fillcolor=white,
    linestyle=solid,
    linecolor=black,
    linewidth=2pt]{#1} 
\fi
}

\newcommand{\SyntheticElementTextFormat}[4]
{
\ifpdf
  \CommonElementTextFormat{#1}{#2}{\OutlineText{\huge #3}}{#4}
\else
  \CommonElementTextFormat{#1}{#2}{\OutlineText{\Large #3}}{#4}
\fi
}


\begin{tikzpicture}[font=\sffamily, scale = 0.32, transform shape]

  \tikzstyle{ElementFill} = [fill=white]
  \tikzstyle{ParaFill} = [fill=gray!50]
  \tikzstyle{FerroFill} = [fill=blue!45]
  \tikzstyle{SpinFill} = [fill=red!65]
  \tikzstyle{SCFill} = [fill=teal!65]
  \tikzstyle{AFMFill} = [fill=yellow!55]

  \tikzstyle{Element} = [draw=black, ElementFill,
    minimum width=2.55cm, minimum height=2.55cm, node distance=2.55cm]
  \tikzstyle{Para} = [Element, ParaFill]
  \tikzstyle{Ferro} = [Element, FerroFill]
  \tikzstyle{Spin} = [Element, SpinFill]
  \tikzstyle{SC} = [Element, SCFill]
  \tikzstyle{AFM} = [Element, AFMFill]
  \tikzstyle{PeriodLabel} = [font={\rmfamily\LARGE}, node distance=2.0cm]
  \tikzstyle{GroupLabel} = [font={\rmfamily\LARGE}, minimum width=2.5cm, node distance=2.0cm]
  \tikzstyle{TitleLabel} = [font={\rmfamily\Huge\bfseries}]

  \node[name=H, Element] {\NaturalElementTextFormat{1}{1.0079}{H}{Hydrogen}};
  \node[name=Li, below of=H, Element] {\NaturalElementTextFormat{3}{6.941}{Li}{Lithium}};
  \node[name=Na, below of=Li, Element] {\NaturalElementTextFormat{11}{22.990}{Na}{Sodium}};
  \node[name=K, below of=Na, Element] {\NaturalElementTextFormat{19}{39.098}{K}{Potassium}};
  \node[name=Rb, below of=K, Element] {\NaturalElementTextFormat{37}{85.468}{Rb}{Rubidium}};
  \node[name=Cs, below of=Rb, Element] {\NaturalElementTextFormat{55}{132.91}{Cs}{Caesium}};
  \node[name=Fr, below of=Cs, Element] {\NaturalElementTextFormat{87}{223}{Fr}{Francium}};

  \node[name=Be, right of=Li, Element] {\NaturalElementTextFormat{4}{9.0122}{Be}{Beryllium}};
  \node[name=Mg, below of=Be, Element] {\NaturalElementTextFormat{12}{24.305}{Mg}{Magnesium}};
  \node[name=Ca, below of=Mg, Element] {\NaturalElementTextFormat{20}{40.078}{Ca}{Calcium}};
  \node[name=Sr, below of=Ca, Element] {\NaturalElementTextFormat{38}{87.62}{Sr}{Strontium}};
  \node[name=Ba, below of=Sr, Element] {\NaturalElementTextFormat{56}{137.33}{Ba}{Barium}};
  \node[name=Ra, below of=Ba, Element] {\NaturalElementTextFormat{88}{226}{Ra}{Radium}};

  \node[name=Sc, right of=Ca, Element] {\NaturalElementTextFormat{21}{44.956}{Sc}{Scandium}};
  \node[name=Y, below of=Sc, Element] {\NaturalElementTextFormat{39}{88.906}{Y}{Yttrium}};
  \node[name=LaLu, below of=Y, Element] {\NaturalElementTextFormat{57-71}{}{La-Lu}{Lanthanide}};
  \node[name=AcLr, below of=LaLu, Element] {\NaturalElementTextFormat{89-103}{}{Ac-Lr}{Actinide}};

  \node[name=Ti, right of=Sc, Element] {\NaturalElementTextFormat{22}{47.867}{Ti}{Titanium}};
  \node[name=Zr, below of=Ti, Para] {\ParaTextFormat{}{}{UZr$_2$}{Hex.}};
  \node[name=Hf, below of=Zr, Element] {\NaturalElementTextFormat{72}{178.49}{Hf}{Halfnium}};
  \node[name=Rf, below of=Hf, Element] {\SyntheticElementTextFormat{104}{261}{Rf}{Rutherfordium}};

  \node[name=V, right of=Ti, Element] {\NaturalElementTextFormat{23}{50.942}{V}{Vanadium}};
  \node[name=Nb, below of=V, Element] {\NaturalElementTextFormat{41}{92.906}{Nb}{Niobium}};
  \node[name=Ta, below of=Nb, Element] {\NaturalElementTextFormat{73}{180.95}{Ta}{Tantalum}};
  \node[name=Db, below of=Ta, Element] {\SyntheticElementTextFormat{105}{262}{Db}{Dubnium}};

  \node[name=Cr, right of=V, Element] {\NaturalElementTextFormat{24}{51.996}{Cr}{Chromium}};
  \node[name=Mo, below of=Cr, Element] {\NaturalElementTextFormat{42}{95.94}{Mo}{Molybdenum}};
  \node[name=W, below of=Mo, Element] {\NaturalElementTextFormat{74}{183.84}{W}{Tungsten}};
  \node[name=Sg, below of=W, Element] {\SyntheticElementTextFormat{106}{266}{Sg}{Seaborgium}};

  \node[name=Mn, right of=Cr, AFM] {\MagTextFormat{}{260 K}{UMn$_2$}{Cubic~\cite{UMn2}}};
  \node[name=Tc, below of=Mn, Element] {\NaturalElementTextFormat{43}{96}{Tc}{Technetium}};
  \node[name=Re, below of=Tc, Para] {\ParaTextFormat{}{}{URe$_2$}{Orth.~\cite{URe2}}};
  \node[name=Bh, below of=Re, Element] {\SyntheticElementTextFormat{107}{264}{Bh}{Bohrium}};

  \node[name=Fe, right of=Mn, Ferro] {\MagTextFormat{}{165 K}{UFe$_2$}{Cubic~\cite{UFe2}}};
  \node[name=Ru, below of=Fe, Element] {\NaturalElementTextFormat{44}{101.07}{Ru}{Ruthenium}};
  \node[name=Os, below of=Ru, AFM] {\MagTextFormat{}{41 K}{UOs$_2$}{Cubic~\cite{UOs2}}};
  \node[name=Hs, below of=Os, Element] {\SyntheticElementTextFormat{108}{277}{Hs}{Hassium}};

  \node[name=Co, right of=Fe, Para] {\ParaTextFormat{}{}{UCo$_2$}{Cubic~\cite{UCo2}}};
  \node[name=Rh, below of=Co, Element] {\NaturalElementTextFormat{45}{102.91}{Rh}{Rhodium}};
  \node[name=Ir, below of=Rh, Para] {\ParaTextFormat{}{}{UIr$_2$}{Cubic~\cite{UIr2}}};
  \node[name=Mt, below of=Ir, Element] {\SyntheticElementTextFormat{109}{268}{Mt}{Meitnerium}};

  \node[name=Ni, right of=Co, Ferro] {\MagTextFormat{}{21 K}{UNi$_2$}{Hex.~\cite{UNi2}}};
  \node[name=Pd, below of=Ni, Element] {\NaturalElementTextFormat{46}{106.42}{Pd}{Palladium}};
  \node[name=Pt, below of=Pd, Para] {\ParaTextFormat{}{}{UPt$_2$}{Orth.~\cite{UPt2}}};
  \node[name=Ds, below of=Pt, Element] {\SyntheticElementTextFormat{110}{281}{Ds}{Darmstadtium}}; 

  \node[name=Cu, right of=Ni, Element] {\NaturalElementTextFormat{29}{63.546}{Cu}{Copper}};
  \node[name=Ag, below of=Cu, Element] {\NaturalElementTextFormat{47}{107.87}{Ag}{Silver}};
  \node[name=Au, below of=Ag, Para] {\ParaTextFormat{}{}{UAu$_2$*}{Hex.~\cite{UAu2}}};
  \node[name=Rg, below of=Au, Element] {\SyntheticElementTextFormat{111}{280}{Rg}{Roentgenium}};

  \node[name=Zn, right of=Cu, Element] {\NaturalElementTextFormat{30}{65.39}{Zn}{Zinc}};
  \node[name=Cd, below of=Zn, Element] {\NaturalElementTextFormat{48}{112.41}{Cd}{Cadmium}};
  \node[name=Hg, below of=Cd, AFM] {\MagTextFormat{}{80 K}{UHg$_2$}{Cubic~\cite{UHg2}}};
  \node[name=Cn, below of=Hg, Element] {\SyntheticElementTextFormat{112}{285}{Cn}{Copernicium}};

  \node[name=Ga, right of=Zn, Ferro] {\MagTextFormat{}{124 K}{UGa$_2$}{Hex.~\cite{UGa2}}};
  \node[name=Al, above of=Ga, Spin] {\MagTextFormat{}{26 K}{UAl$_2$}{Cubic~\cite{UAl2}}};
  \node[name=B, above of=Al, Para]{\ParaTextFormat{}{}{UB$_2$}{Hex.~\cite{UB2}}};
  \node[name=In, below of=Ga, Element] {\NaturalElementTextFormat{49}{114.82}{In}{Indium}};
  \node[name=Tl, below of=In, Element] {\NaturalElementTextFormat{81}{204.38}{Tl}{Thallium}};
  \node[name=Nh, below of=Tl, Element] {\SyntheticElementTextFormat{113}{284}{Nh}{Nihonium}};

  \node[name=C, right of=B, Para] {\ParaTextFormat{}{}{UC$_2$}{Cubic~\cite{UC2}}};
  \node[name=Si, below of=C, Para]  {\ParaTextFormat{}{}{USi$_2$}{Hex.~\cite{USi2}}};
  \node[name=Ge, below of=Si, Ferro] {\MagTextFormat{}{52 K}{UGe$_2$}{Orth.~\cite{UGe2}}};
  \node[name=Sn, below of=Ge, AFM] {\MagTextFormat{}{75 K}{USn$_2$}{Orth.~\cite{USn2}}};
  \node[name=Pb, below of=Sn, Element] {\NaturalElementTextFormat{82}{207.2}{Pb}{Lead}};
  \node[name=Fl, below of=Pb, Element] {\SyntheticElementTextFormat{114}{289}{Fl}{Flerovium}};

  \node[name=N, right of=C, Element] {\NaturalElementTextFormat{7}{14.007}{N}{Nitrogen}};
  \node[name=P, below of=N, AFM] {\MagTextFormat{}{203 K}{UP$_2$}{Tet.~\cite{UP-Bi}}};
  \node[name=As, below of=P, AFM] {\MagTextFormat{}{273 K}{UAs$_2$}{Tet.~\cite{UP-Bi}}};
  \node[name=Sb, below of=As, AFM] {\MagTextFormat{}{203 K}{USb$_2$}{Tet.~\cite{UP-Bi}}};
  \node[name=Bi, below of=Sb, AFM] {\MagTextFormat{}{194 K}{UBi$_2$}{Tet.~\cite{UP-Bi}}};
  \node[name=Mc, below of=Bi, Element] {\SyntheticElementTextFormat{115}{288}{Mc}{Moscovium}};

  \node[name=O, right of=N, AFM] {\MagTextFormat{}{31 K}{UO$_2$}{Cubic~\cite{UO2}}};
  \node[name=S, below of=O, Para] {\ParaTextFormat{}{}{$\beta$-US$_2$}{Orth.~\cite{US2}}};
  \node[name=Se, below of=S, Ferro] {\MagTextFormat{}{14 K}{$\beta$-USe$_2$}{Orth.~\cite{USe2}}};
  \node[name=Te, below of=Se, SC] {\MagTextFormat{}{1.6 K}{UTe$_2$}{Orth.~\cite{UTe2}}};
  \node[name=Po, below of=Te, Element] {\NaturalElementTextFormat{84}{209}{Po}{Polonium}};
  \node[name=Lv, below of=Po, Element] {\SyntheticElementTextFormat{116}{293}{Lv}{Livermorium}};

  \node[name=F, right of=O, Element] {\NaturalElementTextFormat{9}{18.998}{F}{Flourine}};
  \node[name=Cl, below of=F, Element] {\NaturalElementTextFormat{17}{35.453}{Cl}{Chlorine}};
  \node[name=Br, below of=Cl, Element] {\NaturalElementTextFormat{35}{79.904}{Br}{Bromine}};
  \node[name=I, below of=Br, Element] {\NaturalElementTextFormat{53}{126.9}{I}{Iodine}};
  \node[name=At, below of=I, Element] {\NaturalElementTextFormat{85}{210}{At}{Astatine}};
  \node[name=Ts, below of=At, Element] {\SyntheticElementTextFormat{117}{292}{Ts}{Tennessine}}; 

  \node[name=Ne, right of=F, Element] {\NaturalElementTextFormat{10}{20.180}{Ne}{Neon}};
  \node[name=He, above of=Ne, Element] {\NaturalElementTextFormat{2}{4.0025}{He}{Helium}};
  \node[name=Ar, below of=Ne, Element] {\NaturalElementTextFormat{18}{39.948}{Ar}{Argon}};
  \node[name=Kr, below of=Ar, Element] {\NaturalElementTextFormat{36}{83.8}{Kr}{Krypton}};
  \node[name=Xe, below of=Kr, Element] {\NaturalElementTextFormat{54}{131.29}{Xe}{Xenon}};
  \node[name=Rn, below of=Xe, Element] {\NaturalElementTextFormat{86}{222}{Rn}{Radon}};
  \node[name=Og, below of=Rn, Element] {\SyntheticElementTextFormat{118}{294}{Og}{Oganesson}};

  \node[name=Period1, left of=H, PeriodLabel] {1};
  \node[name=Period2, left of=Li, PeriodLabel] {2};
  \node[name=Period3, left of=Na, PeriodLabel] {3}; 
  \node[name=Period4, left of=K, PeriodLabel] {4}; 
  \node[name=Period5, left of=Rb, PeriodLabel] {5};
  \node[name=Period6, left of=Cs, PeriodLabel] {6};
  \node[name=Period7, left of=Fr, PeriodLabel] {7};

  \node[name=Group1, above of=H, GroupLabel] {1 \hfill IA};
  \node[name=Group2, above of=Be, GroupLabel] {2 \hfill IIA};
  \node[name=Group3, above of=Sc, GroupLabel] {3 \hfill IIIA};
  \node[name=Group4, above of=Ti, GroupLabel] {4 \hfill IVB};
  \node[name=Group5, above of=V, GroupLabel] {5 \hfill VB};
  \node[name=Group6, above of=Cr, GroupLabel] {6 \hfill VIB};
  \node[name=Group7, above of=Mn, GroupLabel] {7 \hfill VIIB};
  \node[name=Group8, above of=Fe, GroupLabel] {8 \hfill VIIIB};
  \node[name=Group9, above of=Co, GroupLabel] {9 \hfill VIIIB};
  \node[name=Group10, above of=Ni, GroupLabel] {10 \hfill VIIIB};
  \node[name=Group11, above of=Cu, GroupLabel] {11 \hfill IB};
  \node[name=Group12, above of=Zn, GroupLabel] {12 \hfill IIB};
  \node[name=Group13, above of=B, GroupLabel] {13 \hfill IIIA};
  \node[name=Group14, above of=C, GroupLabel] {14 \hfill IVA};
  \node[name=Group15, above of=N, GroupLabel] {15 \hfill VA};
  \node[name=Group16, above of=O, GroupLabel] {16 \hfill VIA};
  \node[name=Group17, above of=F, GroupLabel] {17 \hfill VIIA};
  \node[name=Group18, above of=He, GroupLabel] {18 \hfill VIIIA};

  \node at ($(H.west -| Fe.north)+ (-3em,2em)$) [name=diagramTitle, TitleLabel]
    {Structural and Magnetic Properties of U$X_2$};
  
   \draw[black, FerroFill] ($(Li.west -| Ti.north) + (0em,-3em)$)
    rectangle +(3em, 3em) node[right, xshift=3ex, yshift=-3.5ex]{\Huge $T_C$};
   \draw[black, AFMFill] ($(Li.west -| Ti.north) + (12em,-3em)$)
    rectangle +(3em, 3em) node[right, xshift=3ex, yshift=-3.5ex]{\Huge $T_N$};
     \draw[black, SCFill] ($(Li.west -| Ti.north) + (24em,-3em)$)
    rectangle +(3em, 3em) node[right, xshift=3ex, yshift=-3.5ex]{\Huge $T_{sc}$};
    \draw[black, SpinFill] ($(Li.west -| Ti.north) + (36em,-3em)$)
    rectangle +(3em, 3em) node[right, xshift=3ex, yshift=-3.5ex]{\Huge $T_{sf}$};
     \draw[black, ParaFill] ($(Li.west -| Ti.north) + (48em,-3em)$)
    rectangle +(3em, 3em) node[right, xshift=3ex, yshift=-3.5ex]{\Huge Para};

\end{tikzpicture}
\caption{(Color online) Summary of the magnetic properties, crystal structures, and magnetic ordering temperatures of U$X_2$ system. Color 45\% blue is used to mark ferromagnetic ordering, 55\% yellow for AFM order, 65\% teal for superconductivity, 65\% red for spin fluctuation, and 50\% gray for paramagnetism. It is worth noting that in the case of UAu$_2$ a serious controversy about its magnetic properties still remains~\cite{UAu2}.}
\end{figure*}

In this paper, we focus on the physical properties of $\delta$-UZr$_2$, for the first time, measured from 1.8 to 300 K and under magnetic fields up to 8 T. We show that all results obtained strongly point to the presence of delocalized 5$f$-electrons in this material. Furthermore, the transport properties show characteristics typical of disordered metallic systems. We also performed electronic structure calculations and compare the results to experimental measurements.

\section{Experimental details}

Polycrystalline samples with nominal compositions $\delta$-UZr$_2$ were synthesized by arc melting stoichiometric amounts of the elements in a Zr-gettered ultra-pure argon atmosphere~\cite{TK}. The samples were examined by TEM and x-ray diffraction measurements. The crystal structure is shown to be hexagonal with the AlB$_2$ structure type, S.G. P6/mmm with the lattice parameters $a$ = 5.036 $\rm{\AA}$ and $c$ = 3.094 $\rm{\AA}$. The values of the lattice parameters are very close to those derived previously~\cite{deltaUZr2}. Also, no other diffraction peaks than expected for AlB$_2$ of the structure were observed. Magnetization, resistivity, heat capacity, Seebeck effect, and thermal conductivity measurements have been performed using a Quantum Design PPMS DynaCool-9 system equipped with a 9 T superconducting magnet with VSM, ETO, HCP and TTO options. Density Functional Theory (DFT) calculations within the Local Density Approximation (LDA)~\cite{Perdew19815048} were performed using the Projector Augmented Wave (PAW) method~\cite{Blochl199417953,Kresse19991758}, as implemented in the VASP code~\cite{Kresse1993558,Kresse199414251,Kresse199615,Kresse199611169}.  A plane wave basis with a kinetic energy cutoff of 520 eV was employed. We used a $\Gamma$-centered \textbf{k}-point mesh of 20$\times$20$\times$20. The crystal structure was relaxed, yielding lattice parameters of $\textbf{a}_1 = (a,0,0)$, $\textbf{a}_2 = a/2(-1, \sqrt 3, 0)$, and $\textbf{a}_3 = (0,0,c)$, where $a$ = 5.12036\AA, $c$ = 2.78937\AA.

\section{Results and discussion}

\begin{figure}[htbp]
\includegraphics[width= 0.5 \textwidth]{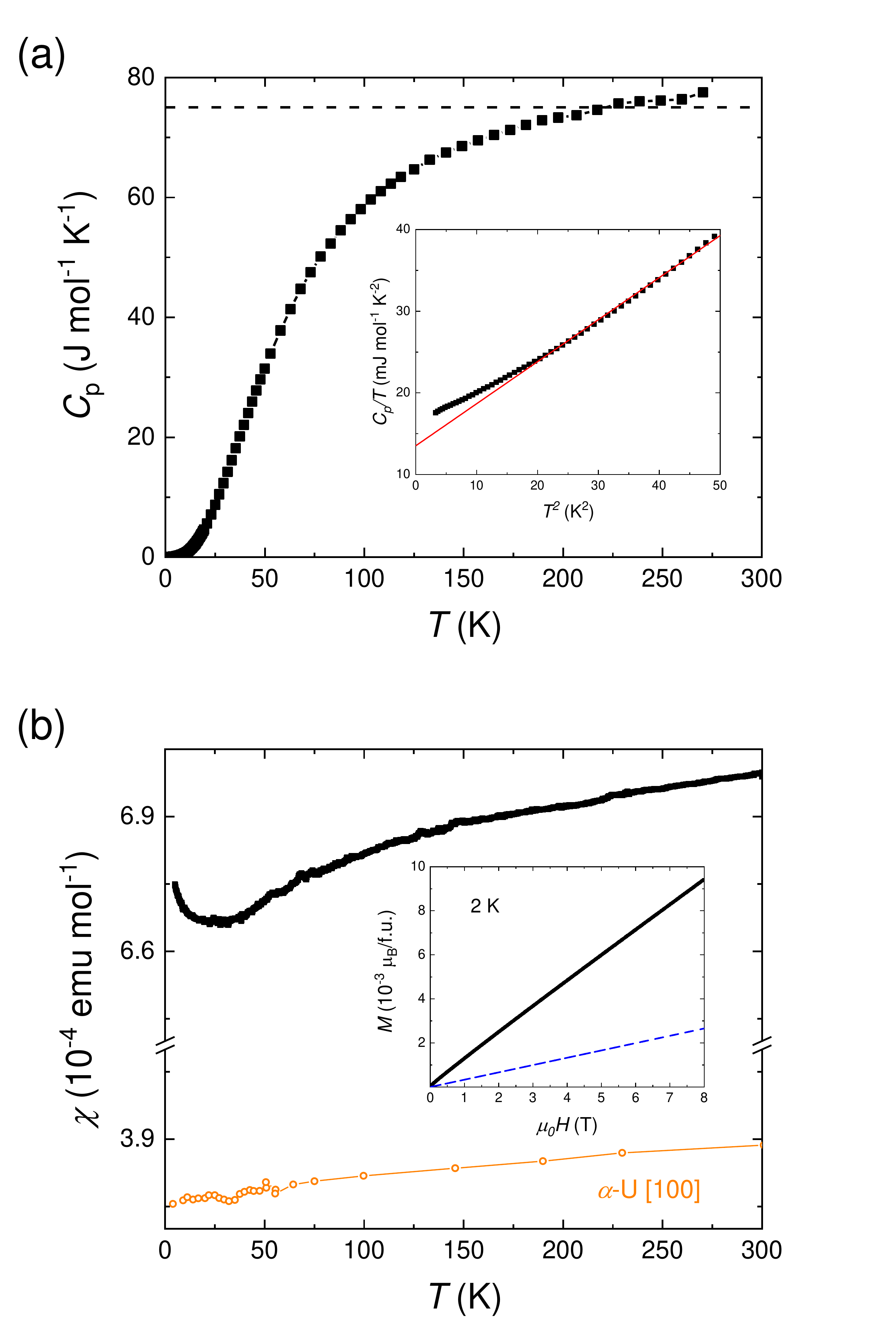}
\caption{(Color online) (a) Temperature dependence of heat capacity of $\delta$-UZr$_2$ from 1.8 to 270 K. The dashed line is the Dulong-Petit limit. Inset shows $C_p/T$ versus $T^2$ below 7 K, the red line is a linear fit. (b) Temperature dependence of the magnetic susceptibility measured at $\mu_0H$ = 5 T and in the temperature range 5 - 300 K. Orange dots represent the $\chi_P(T)$ of $\alpha$-U single crystal measured along [100], data taken from Ref.~\cite{alpha}. Inset: magnetic field dependence of magnetization at $T$ = 2 K. The blue dashed line shows the calculated zero-temperature Pauli magnetization.}\label{t}
\end{figure}

The temperature dependence of the heat capacity $C_p(T)$ of $\delta$-UZr$_2$ measured from 1.8 to 270 K is shown in Fig.~\ref{t}(a). At 270 K, the value of $C_p$ is close to 77.5 J mol$^{-1}$ K$^{-1}$. This value is slightly higher than the theoretical Dulong-Petit limit 3$nR$ = 74.8 J mol$^{-1}$ K$^{-1}$, where $n$ is the number of atoms per formula unit (f.u.) and $R$ is the gas constant. The inset shows the low-temperature heat capacity plotted as $C_p/T$ vs. $T^2$. The red line is a fit of $C_p = \gamma T + \beta T^3$, where $\gamma$ is the Sommerfeld coefficient that is proportional to the electronic DOS, and $\beta$ is a term related to the Debye temperature. A small deviation from the fit occurs below 4 K, which might indicate the presence of some additional low-energy excitations. The $\gamma$ value obtained from the fit is 13.5 mJ mol$^{-1}$ K$^{-2}$. An estimation of the electronic heat capacity ($C_{el} = \gamma T$) at 270 K gives 3.6 J mol$^{-1}$ K$^{-1}$, which is close to the deviation 2.7~J~mol$^{-1}$ K$^{-1}$ observed at 270 K. This indicates that the 5$f$-electrons in the $\delta$ phase are only weakly correlated. The electronic DOS at the Fermi energy $E_F$ calculated by expression $N(E_F) = \frac{3\gamma}{\pi^2k{_B^2}N_A}$ is about 5.7 states/(eV f.u.), where $k_B$ is Boltzmann constant and $N_A$ represents the Avogadro number. The Debye temperature of $\delta$-UZr$_2$ can be derived by the formula $\Theta_D = (\frac{12nR\pi^4}{5\beta})^{\frac{1}{3}}$ and equals to 225~K. 

Figure~\ref{t}(b) shows the temperature dependence of the magnetic susceptibility $\chi(T)$ of $\delta$-UZr$_2$, measured from 5 to 300 K in magnetic field of 5 T. The $\chi(T)$ shows a weak temperature dependence with no sign of magnetic phase transitions. Below 25 K an upturn is present, presumably due to the existence of very small amounts of paramagnetic impurities in the samples (most probably lanthanides), which follows the Curie-Weiss law. For comparison (marked by orange circles), we have also included the temperature dependence of the magnetic susceptibility $\chi_P(T)$ of an $\alpha$-U single crystal along [100] (extracted from Ref.~\cite{alpha}). As can be seen, the magnetic susceptibility of $\alpha$-U is of the Pauli-type, which shows a very little temperature dependence with a small decrease as the temperature decreases across the whole temperature range. Ross and Lam suggested that the change of $\chi_P(T)$ might be due to changes in the relative positions of the Brillouin zones and the Fermi surface as the sample contracts anisotropically with decreasing temperature~\cite{alpha}. The overall magnitude and temperature dependence of the magnetic susceptibility of $\delta$-UZr$_2$ is similar to that of $\alpha$-U metal. In addition, the magnetic susceptibility of $\delta$-UZr$_2$ is larger than that of $\alpha$-U. This might indicate the presence of a slightly larger density of states (DOS) in $\delta$-UZr$_2$ than in $\alpha$-U. The inset displays the magnetic field dependence of magnetization $M(H)$ measured at 2 K. The magnetic moment induced at 8 T is only $\sim$ 0.01 $\mu_B$, suggesting delocalized 5$f$-electrons. Take into account $N(E_F)$ obtained above and using the free electron Fermi gas model, the zero-temperature Pauli magnetization could be calculated with $M_P(H) = \mu^2_BN(E_F)\mu_0H$, where $\mu_0$ is the vacuum permeability. As displayed by the blue dashed line, the so-obtained $M_P(H)$ is compared to the measured magnetization of $\delta$-UZr$_2$. The underestimation of $M_P(H)$ might be related to the presence of the small amount of paramagnetic impurities that are not taken into account in this analysis.

\begin{figure}[t!]
\includegraphics[width=0.5\textwidth]{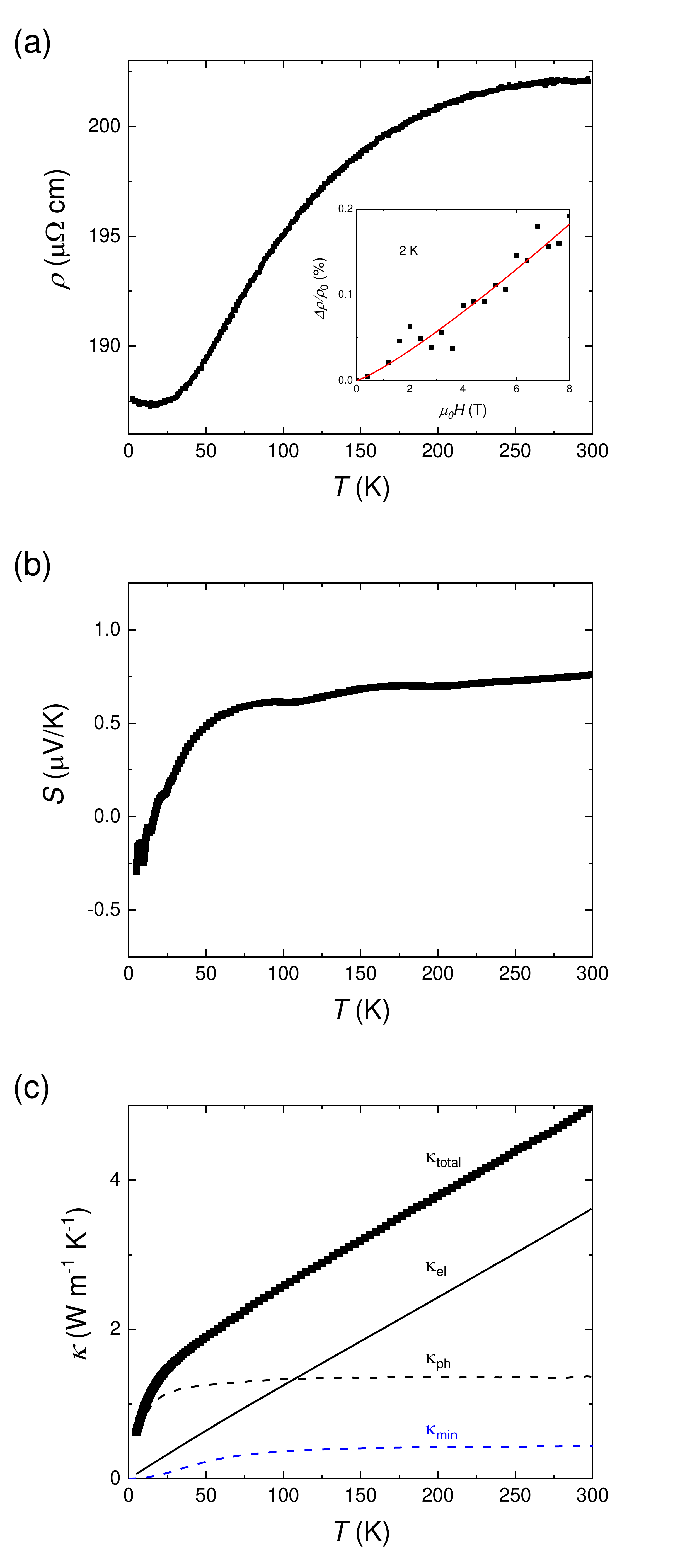}
\caption{(Color online) Transport properties of $\delta$-UZr$_2$. (a) Temperature dependence of resistivity. Inset shows the field dependence of MR at 2 K. Red line shows $H^{1.2}$ behavior. (b) Temperature dependence of Seebeck effect from 5 to 300 K. (c) Temperature dependence of thermal conductivity from 5 to 300 K. The black solid line is the calculated electron contribution $\kappa_{el}$, while the black dashed line shows the subtracted phonon part $\kappa_{ph}$. The blue dashed line shows the estimated minimum phonon contribution to the total thermal conductivity.}\label{tr}
\end{figure}

Temperature dependence of the electrical resistivity $\rho(T)$ of $\delta$-UZr$_{2}$ is shown in Fig.~\ref{tr}(a). The overall shape and magnitude of $\rho(T)$ is typical for uranium intermetallic compounds~\cite{UIr2, U3Si2}. The residual-resistivity ratio (RRR), defined as $\rho(300 K)/\rho(0 K)$ is low and estimated to be $\sim$1.05. This indicates that $\delta$-UZr$_2$ is an electronically disordered system, consistent with disorder in its crystal structure. In general, the low-temperature electron scattering on defects and dislocations results in just a shift in the electrical resistivity towards higher value and hence lowering the RRR value. It will not, however, affect the temperature dependence of resistivity. Besides the $s$-shaped $\rho(T)$~\cite{UIr2}, there is an upturn at low temperatures with the resistivity minimum at 15 K. The low-temperature resistivity upturn, observed in 4$f$- and 5$f$-electron materials is usually associated with Kondo effect~\cite{Kondo4f, UAl2K}. However, in $\delta$-UZr$_2$, this seems to be unlikely because the magnetic susceptibility shows no signatures of localized 5$f$-electrons and the magnetoresistance (MR) is small and positive (see below). Interestingly, the low-temperature resistivity upturn and positive MR have also been observed in ThAsSe~\cite{NMK} and $M$-As-Se ($M$ = Zr, Hf, Th) phases~\cite{NMK2}. Such behavior has been interpreted as a signature of the non-magnetic Kondo effect. However, to draw any firm conclusions on the nature of the low temperature behavior in this material, more studies are required. The inset of Fig.~\ref{tr}(a) shows the magnetic field dependence of MR, defined as $\Delta\rho/\rho_0 = (\rho(H) - \rho_0)/\rho_0$, where $\rho_0$ is the resistivity under zero magnetic field. The value of MR exhibits a very weak field dependence and, at 2 K and 8 T, it reaches only 0.2\%. The red line is a fit of $\Delta\rho/\rho_0 = AH^B$ to the experimental data, where $A$ and $B$ are fitting parameters. The analysis gives $B$ = 1.2 which is smaller than the value of 2 that is observed in normal metals. 

Fig.~\ref{tr}(b) shows the temperature dependence of the Seebeck coefficient $S(T)$. The overall behavior and magnitude of $S(T)$ is characteristic of metallic materials. The positive value of the Seebeck coefficient might indicate that hole-type carriers dominate the electrical and heat transport. In addition, assuming a single-band model and scattering from atomic disorder being dominant at high temperatures, the Fermi energy can be approximated by $E_{F} = \frac{k^2_B\pi^2T}{3eS}$. This gives a value of $E_{F}$ = 9.78 eV being similar to those characterizing simple metals~\cite{Bernard}. The estimated effective carrier concentration is of the order of 10$^{28}$ m$^{-3}$.

The thermal conductivity measured at room temperature is 5 W m$^{-1}$ K$^{-1}$, as shown in Fig.~\ref{tr}(c). In intermetallic samples, the thermal conductivity $\kappa$ is the sum of electron and phonon contributions: $\kappa = \kappa_{el} + \kappa_{ph}$. The solid line shows the temperature dependence of thermal conductivity of electrons, which is calculated by the formula $\kappa_{el}(T)= LT/ \rho(T)$, where $L$ is Lorentz number. After subtraction, the temperature dependence of the phonon contribution is shown as the dotted line. In the context of the presence of the atomic disorder in $\delta$-UZr$_2$ and its impact on thermal transport, it is worthwhile to compare the measured lattice thermal conductivity to the theoretically achievable minimum of the phonon contribution to the total thermal conductivity. If no distinction is made between the transverse and longitudinal acoustic phonon modes, the latter may be expressed by~\cite{equation}:

\begin{equation}\label{kmin}
\kappa_{min}(T) = \left({\frac{3n_v}{4\pi}}\right)^{\frac{1}{3}}\frac{(k_BT)^2}{\hslash\Theta_D}\int_{0}^{\frac{\Theta_D}{T}}\frac{x^3e^x}{\left(e^x-1\right)^2}dx,
\end{equation}
where $x=\hslash\omega/k_{B}T$. In the the above equation, $\omega$ is the phonon frequency, $n_v$ is the number of atoms per unit volume, and $\hslash$ is the reduced Planck constant, respectively. By taking into account $\Theta_{D}$ = 225~K and $n_v$ = 4.4 $\times$10$^{28}$ m$^{-3}$ appropriate for $\delta$-UZr$_2$, the obtained $\kappa_{min}(T)$ curve is shown in Fig.~\ref{tr}(c). 

\begin{figure}[t!]
\includegraphics[width=0.5\textwidth]{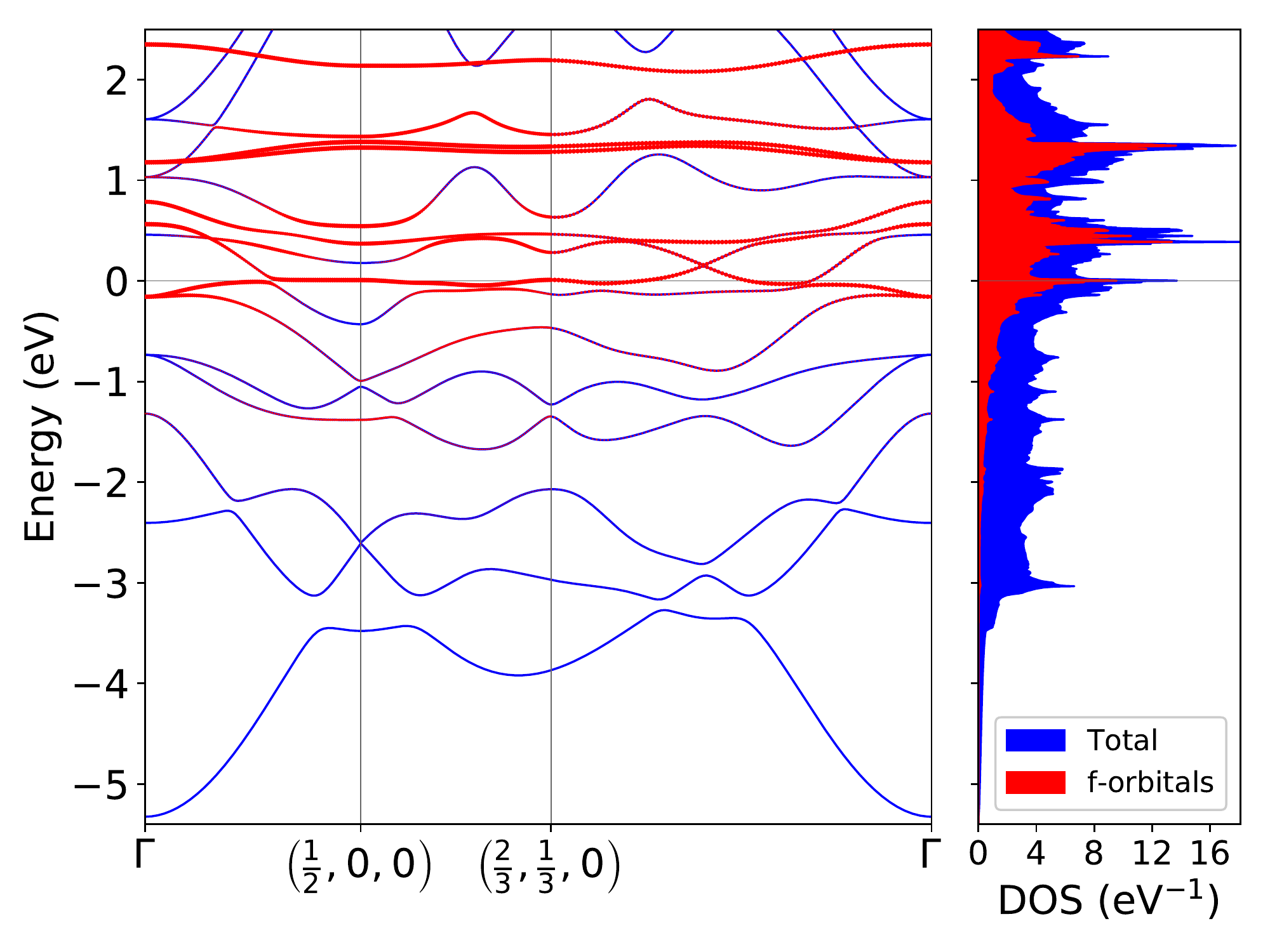}
\caption{(Color online) (Left panel) Electronic band structure of $\delta$-UZr$_2$ computed within DFT (LDA), where the red points indicate the degree of 5$f$-orbital projection for the respective Kohn-Sham Eigenstate. Reciprocal lattice points and distances are given in lattice coordinates. The Fermi energy is at 0, and denoted with a thin line. (Right panel) Corresponding DOS, where blue and red indicate total and 5$f$-projection, respectively.}\label{pdos}
\end{figure}

In order to gain insight into the electronic structure, we perform baseline calculations using DFT within LDA. Since the electronic correlations are not very strong in $\delta$-UZr$_2$ (as indicated by the relatively low value of Sommerfeld coefficient), this approach should give us an overall picture of the electronic structure in this material. We first note that DFT predicts a magnetic instability, in contradiction with experiments. The presence of magnetism within DFT suggests that local correlations will be relevant, and a detailed exploration of this is beyond our current scope. We restrict ourselves to the non-spin-polarized state, and characterize the electronic structure at this level. DFT results for the 5$f$-projected electronic band structure and DOS are provided in Figure~\ref{pdos}, where the width of the red points denotes the degree of 5$f$-projection of the Kohn-Sham Eigenvector. The relatively flat $f$-bands lead to large peaks in the DOS; one of which is nearly at the Fermi energy. The total DOS at the Fermi energy is approximately 12.3 states/(eV f.u.), larger than the experimental value of 5.7 states/(eV f.u.), though allowing magnetism and/or disorder would greatly reduce this value. This key comparison between experimental and theoretical results point to future work, using more sophisticated analysis such as DFT + DMFT (together with disorder effects), to properly capture the nonmagnetic, metallic state in this system.

\section{Conclusions}
In summary, we report on the magnetic, transport, and thermal properties of the $\delta$-phase UZr$_2$, for the first time, measured from 1.8 to 300 K and under magnetic fields up to 8 T. All the results obtained, especially a Pauli type of magnetic susceptibility, small Seebeck and Sommerfeld coefficient strongly point to the presence of delocalized 5$f$-electrons in this material. Transport properties, especially the small RRR value, are indicative of electronic disorder in this metallic system, consistent with its disordered crystal structure. We also performed electronic structure calculations and compare the results to experimentally obtained for the total DOS at the Fermi energy. Although the calculations support the presence of the delocalized 5$f$ states in $\delta$-UZr$_2$, some discrepancies exists, mainly due to the effects of strong electronic correlations that are not sufficiently captured by the LDA.

\section{Acknowledgments}
This work was supported by the US DOE BES Energy Frontier Research Centre "Thermal Energy Transport under Irradiation" (TETI). The electronic structure calculations have been performed using resources of the National Energy Research Scientific Computing Center, a DOE Office of Science User Facility supported by the Office of Science of the U.S. Department of Energy under Contract No. DE-AC02-05CH11231.

\bibliography{UZr2_transport}
\end{document}